\documentclass{ws-procs975x65}
\def\beq{\begin{equation}}
\def\eeq{\end{equation}}
\begin{document}

\title{Tachyonic approach to neutrino dark matter}
\author{Eduard G. Mychelkin$^*$ and Maxim A. Makukov}

\address{Fesenkov Astrophysical Institute, Almaty 050020, Kazakhstan \\
	$^*$E-mail: mychelkin@aphi.kz, edmych@gmail.com}

\begin{abstract}
We apply a new approach based on three relativistic groups (bradyon, tachyon and instanton) forming the `Lorentz groupoid' which allows, in particular, to consider tachyons without introducing imaginary masses and negative energies (related, as known, to violation of causality and unitarity). This leads to effectively scalar conglomerate composed of tachyonic neutrino and antineutrino spinor wave functions as a viable model for stationary dark matter. We also briefly discuss a relevant early non-stationary high-energy stage of the universe evolution.
\end{abstract}


\bodymatter

%%%%%%%%%%%%%%%%% now a standard article style for the most part

\section{Introduction: relation between DE and DM}

\indent For the current stage of the universe evolution dark energy (DE) and dark matter (DM) can be considered as independent fundamental scalar fields. The DE density is $\rho_{de} \simeq \rho_{\Lambda} \simeq 0.7\rho_c \simeq 0.7 \times 10^{-29} $g/cm$^3 \simeq 2.5 \times 10^{-47}$ GeV$^4 \simeq (8 \times 10^{-3} $ eV$)^4$. This value proves to be of the same order of magnitude as the mass of  neutrino per its Compton volume, i.e. its "effective density": $\rho_{\nu} \simeq \alpha m_{\nu}/\lambda_{\nu}^3 =\alpha m_{\nu}^4 \simeq\alpha (8 \times 10^{-3}$ eV$)^4, \alpha \lesssim 1$. This coincidence indicates that, in principle, neutrinos might represent perturbances in the DE background (e.g. shock-vortex solitons). For DE model we adopt the embedding of the local Papapetrou metric\cite{Papa} into  cosmological Szekeres-type background\cite{Szekeres}, similar to as it was done in Ref. \refcite{Gautreau} :
\begin{equation}
ds^{2} = e^{ - 2\phi} dt^{2} - e^{2\phi} a^2(t)(dr^{2} + r^{2}d\Omega ^{2}),
\end{equation}
where $a^2(t) = e^{-\Lambda(t-t_{0})^2}$ is obtained under the integrability condition $\Lambda = -(2/3)\mu^2$ (see Refs. \refcite{mych09,mychmak2015}) connecting lambda and mass terms of scalar field which in its turn implies that mass scale for DE is $|\mu|=m \approx 10^{-33}$ eV. In our approach the local scalar field in (1) is positively defined, $\phi=\phi(r)>0$, and DE as a whole is identifiable with the neutral superposition $\phi=(\phi^{+}+\phi^{-})/\sqrt{2}$ of primordial quasi-static electric fields \cite{mych09,mychmak2015}. For a central mass $M$ on DE background with $\phi=\phi_{Newton}$ we get:
\begin{equation}
ds^{2} = e^{ - 2GM / r}dt^{2} - e^{2GM /r}e^{-\Lambda(t-t_{0})^2}(dr^{2} + r^{2}d\Omega ^{2}).
\end{equation}
Then stationary DM related to some scalar field $\phi_{dm}(r)$ might be included into $\phi$:
\begin{equation}
\phi=\phi_{Newton}(r)+\phi_{dm}(r).
\end{equation}
\indent Mass and lambda terms of DE are of the same order (mass scale $\approx 10^{-33}$ eV) and so they should enter the Lagrangian (or be neglected) only simultaneously. As for DM, in our approach it proves to be of neutrino mass scale (a fraction of 1 eV). 
\section{Typical problems with tachyons}
We consider approach based on the representation of DM as gravitating tachyon neutrino $\nu\bar\nu$-background. The problem under consideration is how to build DM scalar field from spinor neutrino wave functions and to justify the viability of tachyonity.\\ 
\indent It is well known from experiments on parity violation in weak interactions that all neutrinos are left (and antineutrinos are right). On the other hand, effects of neutrino oscillations hint that neutrinos have nonzero masses and thus cannot travel at the speed of light. If neutrino velocities were less than the speed of light, in some reference frames neutrino helicity would swap to the opposite. As this has never been observed, the conclusion is that neutrino velocity might be (within the accuracy of experiments) greater than the speed of light. In this case tachyonity of neutrinos is not an {\it ad hoc} hypothesis, but the consequence of the chiral invariance. There are independent evidences in favour of the tachyonity of neutrinos as well.\cite{Ehrlich}\

\indent But tachyonity is related to such `eternal' problems as instability of tachyonic modes, negative energies, violation of causality and unitarity, violation of the Pauli principle, problems with the (secondary) quantization, etc.
 
\indent Typically, tachyons arise due to change of sign of the square of mass in the dispersion relation $E^{2}-p^{2}=m^{2}$, so that we get: $E^{2}-p^{2}=(im)^2=-m^{2}$. This leads to description of tachyons as particles with imaginary masses. In this case classification of tachyons on the basis of tachyon representations of Lorentz group does not yield any definite values for spin of tachyons or, at best, leads to infinite spin representations\cite{Bekaert}. As a result, we meet principal difficulties in direct realization of the canonical quantization of tachyons\cite{Perepel}. Unfortunately, this is an unresolvable task today, and so here we retain them unquantized. We cannot also exclude the possibility that tachyons should be treated with another theory (e.g., strings, etc.) beyond QFT.\\ 
\indent In any case, there is only one type of experimentally known particles capable of pretending for tachyonity, and these are neutrinos (of three generations).\\
\indent To this end we consider below a group-theoretical algorithm for description of tachyons which does not require application of imaginary masses and negative energies, and therefore does not lead to violation of causality or unitarity.
\section{Lorentz groupoid}
We suggest preservation of the reality of mass in tachyonic dispersion relation as a crucial step. This means rewriting it as $\tilde{p}^{2}-\tilde{E}^{2}=m^{2},$ which should be reinterpreted as a change of the metric signature to the opposite. However, the change of the signature is impossible within the usual (bradyon) Lorentz group SO(1,3) where the metric tensor is a strict invariant. So, in fact, we transfer to another group which might be termed `tachyon Lorentz group' $\widetilde{SO}(3,1)$ with another parametrization and action on another Minkowski-type space.\\
\indent The mapping between SO(1,3) and $\widetilde{SO}(3,1)$ is implemented by means of the inversion of squared velocities algorithm\footnote{We apply here the relation between \emph{squared} velocities or modules of \textbf{v} and \textbf{u} to emphasize that they belong to different spacetimes, and therefore cannot be used to form scalar products of type \textbf{uv} (as done, e.g., in Ref.~\refcite{Recami}) or in velocity addition formulae.}, i.e. superluminal reparametrization of boosts which automatically changes the signature ($c=1$):
\begin{equation}
v^2=1/u^2 \quad \rightleftarrows \quad u^2=1/v^2, \qquad 0\le v<1, \quad 1<u<\infty.
\label{inversion}
\end{equation}
Thus, one gets a one-to-one correspondence (with the exclusion of light cones $v=1$ and $u=1$) between bradyons and (tilded) tachyons:
\begin{equation}
E^{2} (v)-p^{2} (v)=m^{2} \quad \rightleftarrows \quad \tilde{p}^{2} (u)-\tilde{E}^{2}(u)=m^{2},
\end{equation}
where, as usual, bradyon and tachyon energies are, correspondingly: 
\begin{equation}
E(v)=m/\sqrt{(1-v^2)}, \qquad \quad \tilde{E}(u)=m/\sqrt{({{u}^{2}}-1)}.
\end{equation}
\indent Similarly, by changing the sign in square-velocity inversion algorithm,
\begin{equation}
u^2=-1/w_1^2, \quad v^2=-1/w_2^2,  \qquad 0\le w_1<1, \quad 1<w_2<\infty,
\label{inversion2}
\end{equation}
 one gets correspondence (excluding the point $w_1=w_2=1$):
\begin{equation}
\tilde{p}^{2} (u)-\tilde{E}^{2} (u)=m^{2} \quad \rightleftarrows \quad \hat{E}^{2} (w_1)+\hat{p}^{2} (w_1)=m^{2},
\end{equation}
\begin{equation}
E^{2} (v)-p^{2} (v)=m^{2} \quad \rightleftarrows \quad \hat{p}^{2} (w_2)+\hat{E}^{2} (w_2)=m^{2},
\end{equation}
Thus, we have come over to the two adjacent subgroups -- subluminal $SO(4)_1$ and superluminal $SO(4)_2$ -- of the full Euclidean (instanton) Lorentz group $SO(4)$ acting on the Euclidean space-time.\cite{Ramond} The dispersion relations for these subgroups are connected through the mapping $w_1^2=1/w_2^2$ similar to (5):
\begin{equation}
\hat{E}^{2} (w_1)+\hat{p}^{2} (w_1)=m^{2} \quad \rightleftarrows \quad \hat{p}^{2} (w_2)+\hat{E}^{2} (w_2)=m^{2},
\end{equation}
where $\hat{E}(w_{1,2})=m/\sqrt{(1+{w^2_{1,2}})}$. All these interconnected Lorentz-type groups (bradyon, tachyon and instanton) are combined into the "Lorentz groupoid". Each of them acts on its own tangent 4-space of the corresponding  tangent bundle. 

\section{Tachyon neutrino-antineutrino conglomerate} 
So, square-velocity inversion (4) generates the change of the signature in dispersion relations (5) and simultaneously induces a new tachyon Klein-Gordon-type equation 
$$(\Box+m^2)\Phi=0 \quad \rightarrow \quad (\tilde{\Box}+m^2)\tilde{\Phi}=0,$$
with $\Box=g^{\mu\nu}\partial_\mu\partial_\nu \rightarrow \tilde{\Box}=\tilde{g}^{\mu\nu}\tilde{\partial_\mu}\tilde{\partial_\nu}=\tilde{g}^{\mu\nu}\partial_\mu\partial_\nu$. Similarly, the solutions of the type $\Phi\sim e^{-iEt}$ transform into $\tilde{\Phi}\sim e^{-i\tilde{E}t}$ (see, e.g., Ref.~\refcite{Perepel}). Besides, the change of the metric signature induces new Clifford algebra relations for the Dirac $\gamma$-matrices,
$$\gamma_\mu \gamma_\nu+\gamma_\nu \gamma_\mu=2g_{\mu\nu} \quad \rightarrow \quad \tilde{\gamma}_\mu \tilde{\gamma}_\nu+\tilde{\gamma}_\nu \tilde{\gamma}_\mu=2\tilde{g}_{\mu\nu},$$
which lead immediately to a new tachyon Dirac-type equation:
$$ (i\gamma^\nu\partial_\nu-m)\psi=0 \quad \rightarrow \quad (i\tilde{\gamma}^\nu\partial_\nu-m)\tilde{\psi}=0,$$
where summation in the second case is performed with $\tilde{g}_{\mu\nu}$. 

This should be compared with the known Chodos-Jentschura tachyon Dirac-type equation\cite{Chodos, Jentschura} employing the old metric,  $(i\gamma^\nu\partial_\nu-\gamma^5 m)\psi=0.$ But as this equation depends on gauging of $\gamma$-matrices, we prefer to use another gauge-independent tachyon equation first obtained (in different context) by Dirac himself\cite{DiracPositiveEn}:
$$(i\gamma^\nu\partial_\nu-\Gamma m)\psi=0.$$ Here $\Gamma=\gamma^0\gamma^5$ plays the role of imaginary unit, since $\Gamma^2 = -1_{4\times4}$, which is the same in any gauge. It is remarkable that the latter equation might be split into two independent equations (in terms of the Pauli $\sigma$-matrices and operators of momenta), $\left(p_{0} +\vec{\sigma }\vec{p}+m\right)\psi _{R} =0$ and $\left(p_{0} -\vec{\sigma }\vec{p}-m\right)\psi _{L} =0$, separately for right antineutrinos and left neutrinos (such splitting leads to scalar conglomerate of neutrinos and antineutrinos -- see below). Now, comparing the old and new equations one easily finds the relation between new and old $\gamma$-matrices: $$\tilde{\gamma}^\nu=-\Gamma\gamma^\mu.$$
It may be shown that for the Euclidean and tachyon Lorentz groups the direct Hermitian conjugation is necessary in application to spinors\cite{Ramond}, unlike for the old Lorentz group where only Dirac conjugation should be applied via multiplication by Hermitian matrix $\gamma_0$,  $\bar{\psi }=\psi ^{\dag } \gamma _{0}.$ 
Operating with Hermitian conjugation of wave functions, we obtain mass-terms separately for tachyon neutrinos and antineutrinos: 
$$m\psi^\dagger\psi=m({\psi_L}^\dagger\psi_L+{\psi_R}^\dagger\psi_R)=m(\nu^2+\bar{\nu}^2).$$
As a result, superposition of the squares of free tachyon neutrino $\nu $ and antineutrino $\bar{\nu }$ spinor fields might be represented as a scalar conglomerate\footnote{The term `conglomerate', as opposed to `condensate' (not quite appropriately used in Ref. \refcite{mych04}) is applied here, because tachyons have no definite spin values specified according to the Casimir invariants of the Poincar\'e group \cite{Bekaert} and so at macro-scales can be retained unquantized.}:
\begin{equation}
\Phi =\psi ^{\dag } \psi =\nu ^{2} +\bar{\nu }^{2}, 
\label{congl}
\end{equation}
with $\psi =\nu +i\bar{\nu }$. Such effectively scalar tachyon field $\Phi =\phi_{dm}(r)$  leads to the interpretation of quasi-stationary dark matter phenomenon as primordial gravitating neutrino-antineutrino conglomerate \cite{mych09,mychmak2015}.

\section{Relation of tachyon DM to cosmology}
If practically sterile (especially for low energies) neutrino-antineutrino background is distributed all over the universe, it would produce somewhat denser regions (`smoothed halos') around galaxies and clusters. Gravitating neutrino clouds (`geons') were first considered by Brill and Wheeler\cite{Brill}. However, in massless case they are unstable. Tachyonic neutrinos are massive by definition and they can possess almost stiff and thermodynamically stable equation of state, and thus behave as a quasi-isothermal gravitating medium which provides smoothed logarithmic-type potential\cite{makmych2016} leading to the observed flat rotation curves for galaxies and clusters.\\
\indent Scenario of structure formation might proceed so that the formation of galaxies and clusters first develops according to the usual Jeans instability of baryonic matter, with subsequent increase of the concentration of surrounding massive tachyon neutrino-antineutrino conglomerate providing, in its turn, the necessary follow-up growth of rate for the whole process. It is essential here that there are no direct physical constraints on the value of primordial tachyonic neutrino background density, so it can be considered as a free parameter which might be put to be equal to the observed DM density.\\ 
\indent Whether the secondary neutrinos produced since cosmological temperatures of about a few MeV from lepton annihilation exist in bradyonic or tachyonic state is an open question, and the final conclusions are up to the future experiments.
\section {Conclusion and relation to quantum stage}
We do not apply any unidentifiable \emph{ad hoc} scalar fields or particles to obtain DE and DM. As for effectively scalar DM field, we propose neutrinos (provided that they are tachyons) as the only realistic candidate for DM carriers. Tachyonic neutrinos might, in principle, be generated as vortex perturbations (see Introduction) in the process of collision of primordial DE universes, as described in some models \cite{Alexander}.

Here we have envisaged only quasi-static $\nu\bar{\nu}$-background appropriate for description of the current low-energy state of DM. As for the non-stationary initial hot stage of the universe evolution, we believe it is appropriate to mention that the same gravitating $\nu\bar{\nu}$ conglomerate might, in principle, be regarded as a seed material for scenario similar, e.g.,  to Pervushin's `dilaton fabric' producing intermediate vector bosons.\cite{Pervushin} This would be the case if high-energy colliding radial beams of primordial tachyon neutrinos and antineutrinos in the central domain of super-strong gravitational field could be reprocessed into intermediate vector bosons and leptons, $\nu +\bar{\nu }\to W^{+} +W^{-} $, $\nu +\bar{\nu }\to Z$, $\nu +\bar{\nu }\to e^{+} +e^{-} $ (and maybe also into Higgs-like bosons), with the subsequent evolution close to the standard scenario.

The absence of a consistent quantization scheme for tachyons might be circumvented on the basis of the following consideration. Kinematically bradyons and tachyons are practically indistinguishable at high energies due to very small neutrino masses. For example, for masses of  $\sim 10^{-2}$ eV and energies of order 10 GeV (typical for the OPERA and ICARUS 2012 experiments), the deviation of neutrino velocity from the speed of light is $\lvert \Delta c\rvert/c \approx 10^{-24}$ (whereas the accuracy in ICARUS measurements was only $10^{-6}$). But given the velocity inversion algorithm discussed above, each bradyon has its tachyon counterpart with practically the same energy in the  regime $u\rightarrow c = 1$. Therefore, cross-sections obtained at high energies for bradyon-type (fermion) neutrinos within canonical quantization might, in principle, be applied also for corresponding tachyon neutrinos as a reliable upper bound estimation.

One can expect some pecular features for the tachyon sector. Free neutrinos considered as propagating tachyons with infinite spin representations, strictly speaking, can be neither of Dirac nor, even more so, of Majorana type ($\nu$ and $\bar{\nu}$ are equivalent). It might be argued that, in a sense, tachyonic neutrinos are born (and absorbed) as Dirac's but propagate as Majorana's particles. Next, due to the absence of the state of rest for tachyons probably there is no necessity to distinguish between flavour and mass states. Thus, oscillations of neutrinos might be not spontaneous but result solely from interactions with matter and between themselves. In general, physics of tachyon neutrinos promises to be simpler than that of bradyons.

\end{document}